\documentclass[%
 reprint,
 superscriptaddress,
 amsmath,
 amssymb,
 aps,
maxcitenames=3,
floatfix,
]{revtex4-2}

\usepackage[utf8]{inputenc} 
\usepackage{graphicx}    
\usepackage{dcolumn}     
\usepackage{bm}       
\usepackage[separate-uncertainty = true,multi-part-units=single]{siunitx}
\usepackage{subcaption}
\usepackage{color}
\usepackage{mwe}

\usepackage{amsmath}
\usepackage{comment}
\usepackage[separate-uncertainty = true,multi-part-units=single]{siunitx}

\renewcommand{\vec}[1]{\mathbf{{#1}}}

\newcommand{\vecUnit}[1]{\mathbf{\hat{#1}}}

\renewcommand{\onlinecite}[1]{\citenum{#1}}

\graphicspath{{./}}     

\setcitestyle{super}

\begin{document}

\title{Eigenmodes of fractal drums: A numerical student experiment}

\author{Veronica P. Simonsen}
\affiliation{PoreLab, NTNU -- Norwegian University of Science and Technology, NO-7491 Trondheim, Norway}

\author{Nathan Hale}
\affiliation{Department of Physics, NTNU -- Norwegian University of Science and Technology, NO-7491 Trondheim, Norway}

\author{Ingve Simonsen}
\email{Corresponding author: ingve.simonsen@ntnu.no}
\affiliation{Department of Physics, NTNU -- Norwegian University of Science and Technology, NO-7491 Trondheim, Norway}

\date{\today}

\begin{abstract}
 ``Can one hear the shape of a drum?'' was a question posed (and made famous) by mathematician Mark Kac in the mid-1960s. It addresses whether a deeper connection exists between the resonance modes (eigenmodes) of a drum and its shape. Here we propose a numerical experiment, suitable for advanced undergraduate physics students, on the calculation of the eigenmodes of a square Koch fractal drum, for which experimental results do exist. This exercise is designed to develop the students' understanding of the vibrations of fractal drums, their eigenmodes, and potentially their integrated density of states. The students calculate the lowest order eigenmodes of the fractal drum, visualize these modes, and study their symmetry properties. As an extension, the students may investigate the integrated density of states of the fractal drum and compare their findings to the Weyl-Berry conjecture. 
\end{abstract}

\maketitle



\section{\label{sec:1} Introduction}

It is well known that a large drum has a lower fundamental resonance frequency than a smaller drum. Hence, from the tone that a drum makes, you can potentially say something about its size (the area of the membrane). What now if the area of the drum is the same but we change the shape of the drum? Will this change of shape modify the tones of the drum? In 1966, the Polish mathematician Mark Kac published a seminal and influential paper related to this question under the title ``Can one hear the shape of a drum?''~\cite{Kac1966}.

\begin{figure}[t]
 \centering
 \includegraphics[width=0.38 \columnwidth]{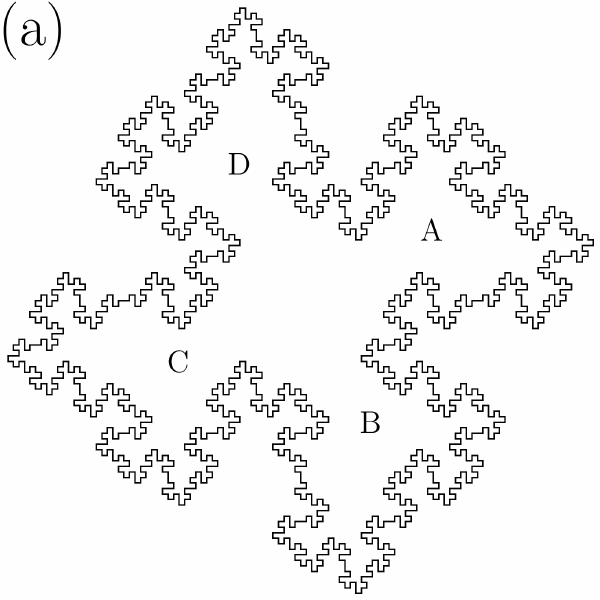} 
 \includegraphics[height=0.4 \columnwidth]{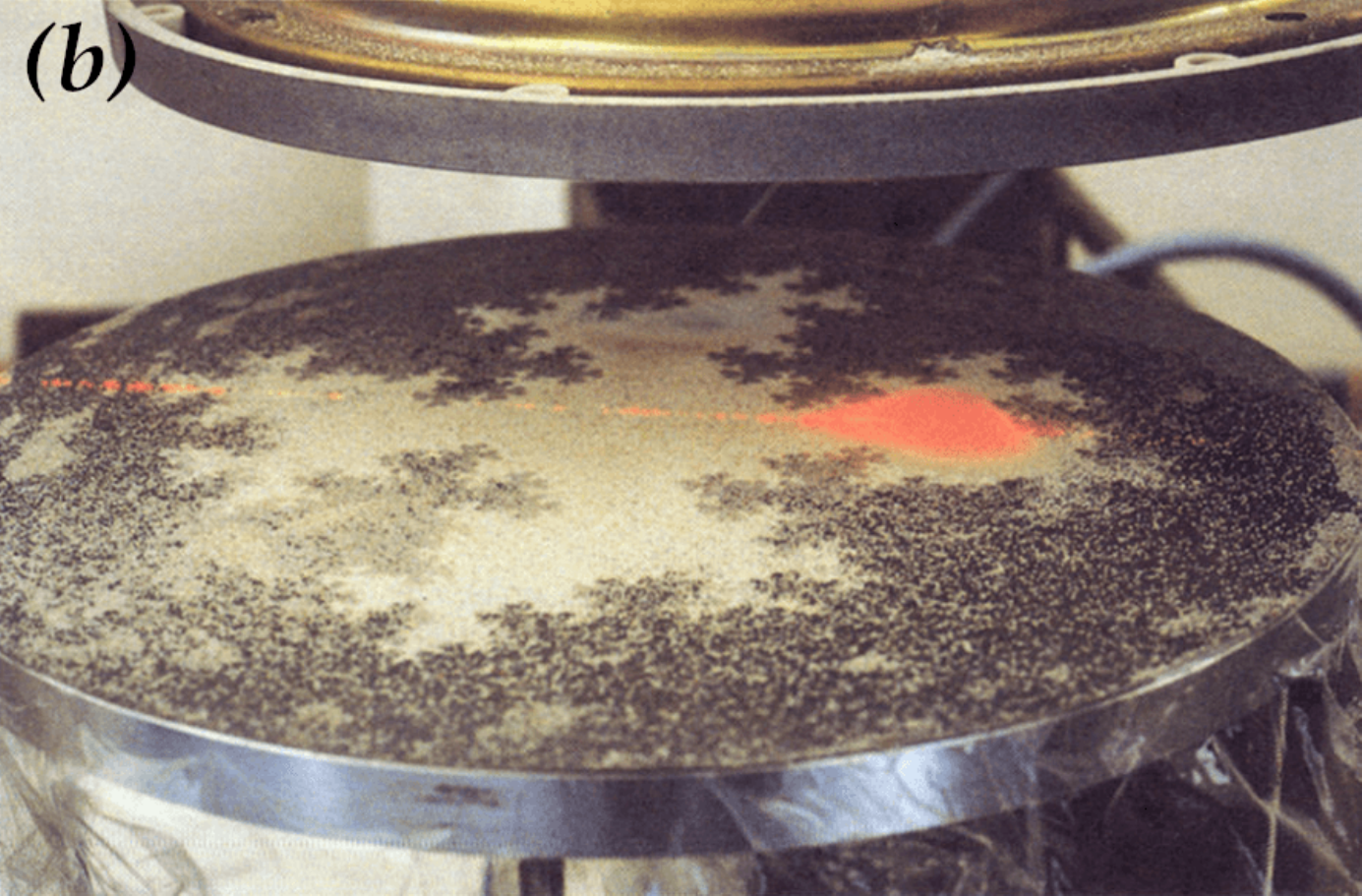}
 \caption{(a) The boundary of the (square Koch) fractal drum~($\ell=3$) that is investigated. The limiting curve has fractal (box-counting) dimension $\ln(8)/\ln(4) = 3/2$. (b) Experimental result of Sapoval~\textit{et al}.~\cite{Sapoval1991} showing localized vibrations (reprinted with permission of APS).}
 \label{Fig:sapoval}
\end{figure}

Shortly after Kac published his famous paper, fractals started to become a topic of interest~\cite{Book:Feder1988}. If the boundary of the drum is \textit{fractal}, and therefore not smooth, what will then happen? In the early 1990s, Sapoval and coworkers conducted a series of elegant experiments to study the modes of fractal drums~\cite{Sapoval1991}. They observed modes \textit{localized} to bounded regions of the drum, labeled $A$, $B$, $C$, and $D$ in Fig.~\ref{Fig:sapoval}(a). In fact, Sapoval~{\textit{et al.}} were able to excite each mode separately. Classic (or non-fractal) drums do not behave this way, as striking any part makes the whole membrane vibrate. Why is the fractal drum so different?

\smallskip
Sapoval~\textit{et al.} showed that the equation governing wave motion has solutions with very large amplitudes at the inward-facing corners of the drum~[Fig.~\ref{Fig:sapoval}]. These large-amplitude regions generate a cascade of large-amplitude vibrations that interfere with one another. This gives rise to dissipation on many scales, so drums with fractal boundaries, hereafter called \textit{fractal drums}, exhibit very strong damping. How does this explain the local vibrations of the fractal drum? The narrow throat connecting region $A$ to the rest of the drum slows a wave traveling from $A$ to $B$~[Fig.~\ref{Fig:sapoval}(a)], and the strong damping absorbs the wave before it can spread. Experimental result for one of these local modes is shown in Fig.~\ref{Fig:sapoval}(b). Any such local modes can be considered a linear combination of the eigenmodes of the system, and the numerical calculation of the possible eigenmodes of the fractal drum is one of the main purposes of the numerical study that we propose here.

In this paper, we introduce a numerical experiment allowing students to study the vibrations of fractal drums, their eigenstates, and potentially their density of states. conjecture. These problems have significant physical applications to the study of porous media, diffusion, wave propagation in fractal media or wave scattering from fractal surfaces.  The tasks are devoted to the numerical calculation of the eigenfrequencies and related eigenmodes of fractal drums. As the perimeter of the fractal drum, we have chosen the so-called square Koch curve, the same structure used in the experiments by Sapoval~{\textit{et al.}}~\cite{Sapoval1991}. The purpose of the numerical experiment that we propose is to enhance students' learning by offering them a means of experimenting with concepts that they may find troublesome in class. Moreover, the experiment is suitable for introductory or upper-level courses and as a modeling exercise in upper-level physics courses.
The experiment can bring enthusiasm to a physics classroom.

\smallskip
The remaining part of this work is organized as follows:
In Sec.~\ref{sec:3} we present the numerical experiment, including its background and the relevant theoretical framework for the fractal drum problem. Then, we provide some implementation details on how to solve the problem and comment on challenges that the students may face in doing so~[Sec.~\ref{Sec:Implementation}]. In Sec.~\ref{sec:4}, we present and discuss the results that were obtained.
Finally, Sec.~\ref{sec:5} presents the conclusions we draw from this work and gives some final remarks.

\section{\label{sec:3} Numerical experiment}

\subsection{Fractal drums}

The problems presented in this work were part of the course \emph{Computational Physics} taught at the Norwegian University of Science and Technology (NTNU). The aim of the fractal drum problem is to numerically calculate the vibrational resonance frequencies of the square Koch drum and obtain the corresponding eigenmodes. This is the same problem that Sapoval and co-workers~\cite{Sapoval1991} studied experimentally in the early 1990s. These authors presented some numerical results for a few eigenmodes of the drum and their results were obtained by a relaxation method (see Ref.~\onlinecite{Sapoval1991} for details). Here, a different numerical approach is used that allows one to obtain all the lower eigenmodes. The physics used in the fractal drum problem, although not explored in this work, extends to applications in the study of porous media, diffusion, wave propagation in fractal media and wave scattering from fractal surfaces.

To state the problem, let $D$ denote the region inside the square Koch drum. The oscillation of the membrane (in $D$) is determined by the wave equation $\nabla^2 u = (1/c^2) \partial^2 u/\partial t^2$ ($c$ is a velocity) subjected to (Dirichlet) boundary condition $u=0$ for all times on the boundary $\partial D$. Here $u(\vec{r},t)$ represents the vertical displacement of the membrane at position $\vec{r}$ in the plane at time $t$. Performing the Fourier transform of the wave equation with respect to time leads to the Helmholtz equation~\cite{Book:Butkov1973,Book:Wong2013}
\begin{subequations}
 \label{eq:Helmholtz}
 \begin{align}
  -\nabla^2 U(\vec{r},\omega) &= \frac{\omega^2}{c^2} U(\vec{r},\omega),   &&\mathrm{in}\; D 
  \label{eq:Helmholtz-A} \\
       U(\vec{r},\omega) &= 0                      &&\mathrm{on}\; \partial D,
  \label{eq:Helmholtz-B} 
 \end{align}
\end{subequations}
where $\omega$ denotes the angular frequency.
Equation~\eqref{eq:Helmholtz-A} states that $\omega^2/c^2$ is an eigenvalue, $\omega$ is the corresponding eigenfrequency, for the negative Laplacian operator [$-\nabla^2$], and the function $U(\vec{r},\omega)$ is  the eigenmode corresponding to the eigenfrequency $\omega$.

A classic approach to solving Eq.~\eqref{eq:Helmholtz} inside $D$ is to use a finite difference approximation to the unknown function $U(\vec{r},\omega)$ in this domain. This is achieved by defining a rectangular grid of lattice constants $h$ in the domain of interest. If $\vec{r}_{mn}=(x_m,y_n)$ represents an arbitrary lattice point in a region of the plane containing the square Koch drum, we let $U(\vec{r}_{mn})=U_{mn}$ denote the vertical displacement of the membrane at this point.

When the standard five-point stencil~\cite{Book:Sauer2012,Book:Abramowitz1964}, defined by the point itself and its four nearest neighbors, is applied to the Laplacian operator that appears on the left-hand side of Eq.~\eqref{eq:Helmholtz-A}, we are led to   
    \begin{multline}
      \label{eq:Helmholtz-FD}
  - \frac{1}{h^2} 
   \left[
   U_{m+1,n} + U_{m-1,n} + U_{m,n+1} + U_{m,n-1} - 4 U_{mn} 
   \right]\\
   =\frac{\omega^2}{c^2} U_{mn}. 
   \end{multline}
The vertical displacement vanishes~[$U_{mn}=0$] for lattice points that are outside, or on the boundary, of the square Koch drum. Hence, it is only the set of displacements $\{U_{mn}\}$ that correspond to lattice points that are inside the square Koch drum that we need to determine. We call these points \emph{internal} lattice points. When Eq.~\eqref{eq:Helmholtz-FD} is applied to all internal lattice points, a set of linear eigenequations is obtained, which determines the eigenfrequencies and the corresponding eigenmodes of the square Koch drum.

\section{Implementation details}
\label{Sec:Implementation}

In this section, we will outline some of the implementation details required to numerically calculate the eigenfrequencies and eigenmodes of the fractal square Koch drum using the finite difference approximation.

\subsection{\label{sec:3.1} Constructing the fractal drum}

\begin{figure}[htp!]
 \centering
 \includegraphics[width=0.55\columnwidth]{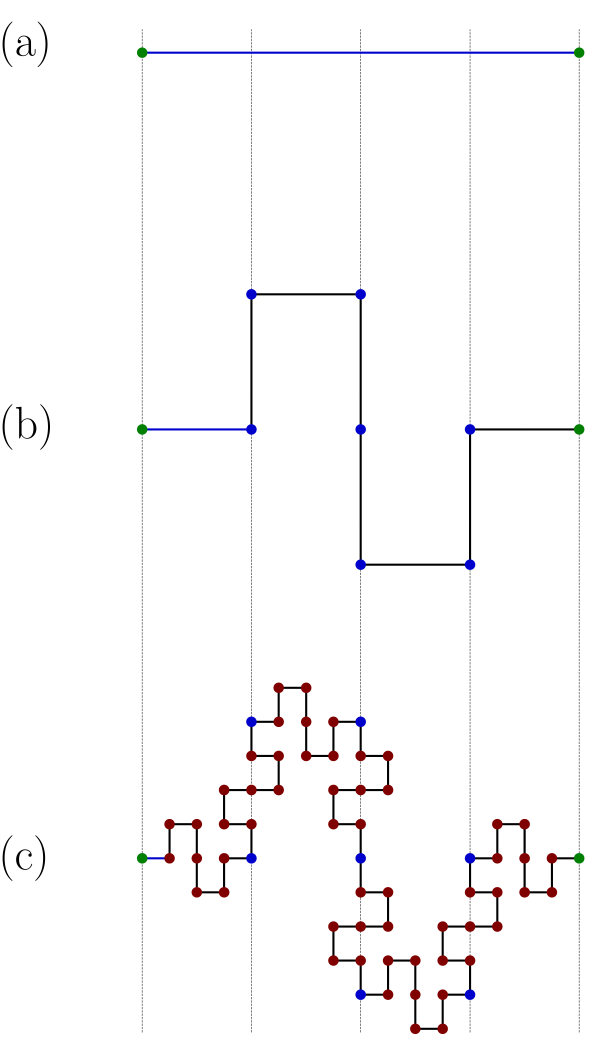}
 \caption{(Color online) The process of constructing the fractal. (a) The initial line segment of length $L$ (level $\ell=0$); (b) the ``\textit{generator}'' of the fractal (level $\ell=1$); (c) the structure at level $\ell=2$ of the construction. The different nodal colors are used to represent the nodes added at each level of the generation process.}
 \label{Fig:fractal-generator}
\end{figure}

The fractal that we will be concerned with is constructed on the basis of the \emph{generator} presented in Fig.~\ref{Fig:fractal-generator}(b). This generator is constructed from an initial ($\ell=0$) line segment of length $L$ [Fig~\ref{Fig:fractal-generator}(a)] by (\textit{i}) dividing it into four equal segments of length $L/4$; (\textit{ii}) raising the 2nd element (from the left) a distance $L/4$ from the base; and (\textit{iii}) lowering the 3rd element a distance $L/4$, while the elements connected to the end points are not moved. It is customary to treat the central vertical part of the generator as two separate line segments instead of one, in order to make each of the $8$ line segments the same length. The structure in Fig.~\ref{Fig:fractal-generator}(b) is the \emph{generator} of the fractal and is represented by generation level $\ell=1$. To obtain the structure at level $\ell=2$, this generator is applied subsequently to each of the line segments of length $L/4$ from the previous generation level. In this way, the $\ell=2$ structure presented in Fig.~\ref{Fig:fractal-generator}(c) is obtained. The structures corresponding to higher levels are generated recursively in the same fashion by applying the generator from Fig~\ref{Fig:fractal-generator}(b) to the smaller-and-smaller line segments from the previous level. In the limit $\ell\rightarrow\infty$, the true fractal structure is obtained; when $\ell$ has a finite value, the structure is said to be a pre-fractal.  Therefore, the drums used in this experiment are technically not a fractal but a pre-fractal.

\begin{figure}[htp!]
 \centering
 \includegraphics[width=0.68\columnwidth]{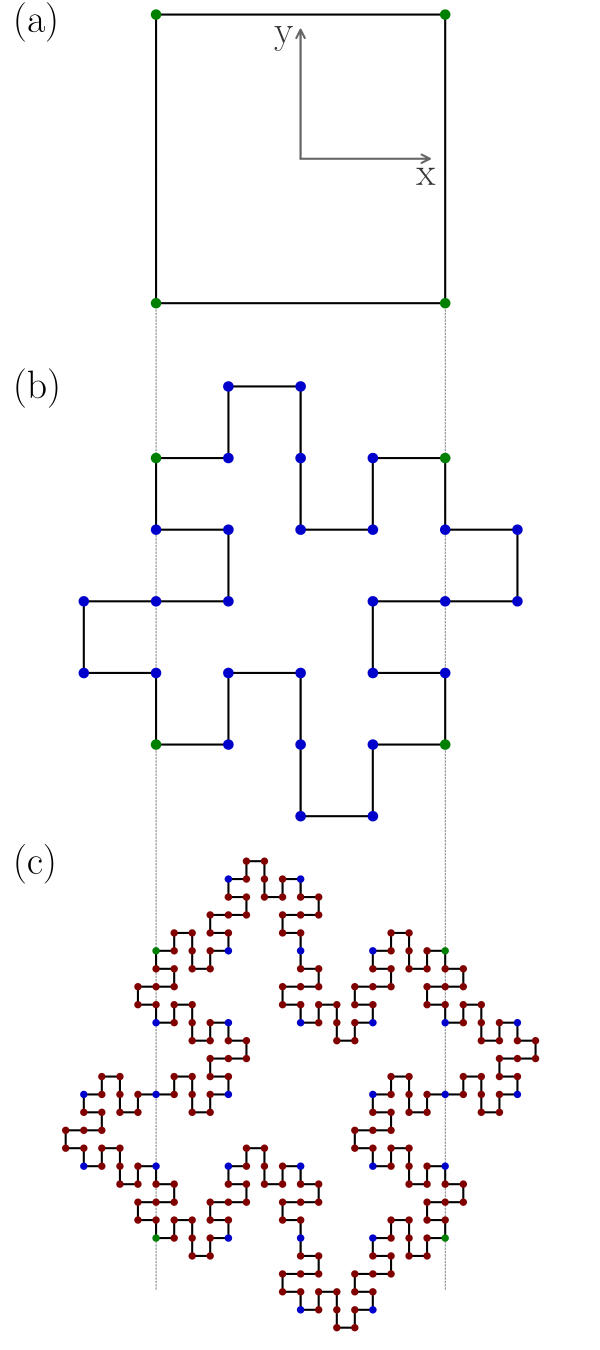} 
 \caption{(Color online) The process of constructing the square Koch fractal. (a) The initial square region (level $\ell=0$) from which the square Koch fractal is generated; (b) level $\ell=1$ of the fractal construction (added points shown in blue); (c) level $\ell=2$ of the construction (added points shown in red). The vertical dashed lines represent the initial width (and height) of the square region from which the square Koch curve is constructed. Notice that the width and height of the structure increase with the generation level $\ell$, but the area inside the curve remains the same at \emph{all} generation levels. The coordinate system that we use is indicated in Panel~(a) and its origin is located at the center of the square.}
 \label{Fig:Koch-fractal}
\end{figure}

The square Koch fractal (of type 2) is generated by starting from a square of sides $L$ [Fig.~\ref{Fig:Koch-fractal}(a)] (level $\ell=0$), and recursively applying the fractal generator from Fig.~\ref{Fig:fractal-generator}(b) to each of its sides. The fractal structure at level $\ell=1$ and $\ell=2$ is obtained and the resulting structures are presented in Figs.~\ref{Fig:Koch-fractal}(b) and ~\ref{Fig:Koch-fractal}(c), respectively. In Fig.~\ref{Fig:Koch-fractal}, the points that are added at each level are presented in different colors. It should be noticed from the way that the structure is generated that the total area inside the structure is $L^2$ and independent of the generation level. Furthermore, the smallest line segment of the structure at level $\ell$ is 
\begin{align}
 \delta_\ell
 &=
  \frac{L}{4^\ell}. 
 \label{eq:discretization-inteval-ell}
\end{align}



\subsection{Discretize}
\label{sec:3.2}

The next step is to introduce a \textit{square} lattice to which all the corners of the square Koch fractal at level $\ell$ belong. For this to be the case, the discretization interval $\delta_\ell$ cannot be independent of the initial width $L$ of the square from which one started the generation (level $\ell=0$). From the structures depicted in Fig.~\ref{Fig:Koch-fractal}, it should be apparent that the widths of the structures grow with generation level $\ell$. From the way the square Koch fractal is generated, one finds that its size (width and height) at level $\ell$ is given as $L_\ell = L + 2\sum_{n=1}^\ell\delta_n$ or as
\begin{align}
 \label{eq:L_over_Leff}
  \frac{L_\ell}{L} 
  &=
  1+2\sum_{n=1}^{\ell} 4^{-n}.
\end{align}
By discretizing a square region of sides $L_\ell$ and using a discretized interval $\delta_\ell = L/4^\ell$, all corners of the square Koch curve at level $\ell$ are guaranteed to fall onto the lattice. If we assume that lattice points coincide with the end points of this square region, a general lattice point is given as 
\begin{subequations}
\begin{align}
  \label{eq:lattice-point}
   \vec{r}_{mn} &= x_m \vecUnit{x} + y_n \vecUnit{y}, 
\end{align}
where the coordinate system used is indicated in Fig.~\ref{Fig:Koch-fractal}(a) and a caret over a vector indicates that it is a unit vector. In writing Eq.~\eqref{eq:lattice-point} we have defined 
\begin{align}
  \label{eq:points-x}
  x_m
  &= -\frac{L_\ell}{2} + (m-1) \delta_\ell 
 \\
 \label{eq:points-y}
  y_n
  &= -\frac{L_\ell}{2} + (n-1) \delta_\ell,
\end{align}
\end{subequations}
with $m=1,2,\ldots,N_\ell+1$ and $n=1,2,\ldots,N_\ell+1$. Here the integer 
\begin{align}
 N_\ell
 &=
  \left\lfloor \frac{L_\ell}{\delta_\ell} \right\rceil
  = \left\lfloor
    4^\ell \left( 1+2\sum_{n=1}^{\ell} 4^{-n} \right)
    \right\rceil,
 \label{eq:No-of-intervals}
\end{align}
denotes the number of line segments (of size $\delta_\ell$) needed to cover the width (or height) of the square region $L_\ell \times L_\ell$ that fully contains the square Koch curve (the symbol $\lfloor\cdot\rceil$ means the nearest integer). The total number of points in the lattice is $(N_\ell+1)^2$ and the fraction of lattice points that are inside the square Koch curve (internal lattice points) can be approximated by the area ratio $(L/L_\ell)^2$~[cf. Eq.~\eqref{eq:L_over_Leff}].

\subsection{Classification of lattice points}
\label{sec:3.3}

To facilitate the implementation of the finite difference expression in Eq.~\eqref{eq:Helmholtz-FD}, it will be beneficial to know which set of lattice points are internal, external, and boundary points for the square Koch curve. To keep track of the classification of the lattice points, we define a square matrix of integers that has a dimension that is identical to the lattice and whose values determine if the lattice point is inside~(positive value), outside~(negative value), or on the boundary (zero value) of the square Koch drum. In the following, we will refer to this matrix as the classification array (or matrix) and it will later be used as a look-up table. Since the corners of the square Koch curve at level $\ell>0$ coincide with some of the lattice points if a lattice constant $h=\delta_\ell$ is used, one can readily identify the boundary points and set the value of the classification array to zero for such points.

It still remains to be determined if the remaining points are inside or outside of the (closed) square Koch curve. The way to do this was \emph{not} specified in the description of the problem that was handed out. Instead, the students were asked to identify and implement at least one method of doing so, and several methods were proposed, implemented, and tested by the students. Here we briefly describe a few such methods.

The (closed) square Koch curve can be seen as a simple polygon since it is defined by its corners. Therefore, our point classification problem is equivalent to the well-known point-in-polygon problem from computer graphics~\cite{Book:Hughes2013,HORMANN2001131,Wikipedia-PIP}. This is an old problem, and numerous algorithms exist to solve it. Here we briefly mention a few that were suggested by students. The \emph{ray casting algorithm}~\cite{Sutherland1974} which keeps track of the number of intersections for a ray (or line) passing from a starting point that is outside (or exterior of) the polygon to the point in question one is investigating; if the number of such intersections is odd, the investigated point is located inside the polygon, if it is even, the point is outside the polygon. In the \emph{winding number algorithm} the investigated point's winding number with respect to the polygon is calculated~\cite{HORMANN2001131}. This number, which is an integer, is zero if the point is outside the polygon, and non-zero if it is inside. The more mathematically inclined students may appreciate that the point-in-polygon problem can be addressed by Cauchy's residue theorem from complex analysis. By defining $z = x + iy$ and letting $z_0$ denote the point of interest, the complex integral $(2\pi i)^{-1}\oint_{\gamma} dz/(z-z_0)$, where $\gamma=\partial D$ is the square Koch curve, will vanish if $z_0$ is outside $\partial D$ and should equal $1$ (the residue of the integrand at $z_0$) if it is inside. By numerically calculating the contour integral it can be determined if a point is inside or outside the square Koch curve. It should be remarked that Cauchy’s residue theorem can be used to define the winding number algorithm since the winding number is just an alternate form of the Cauchy integral given above~\cite{Narasimhan1985}.

\begin{figure}[htp!]
 \centering
 \includegraphics[width=0.99 \columnwidth]{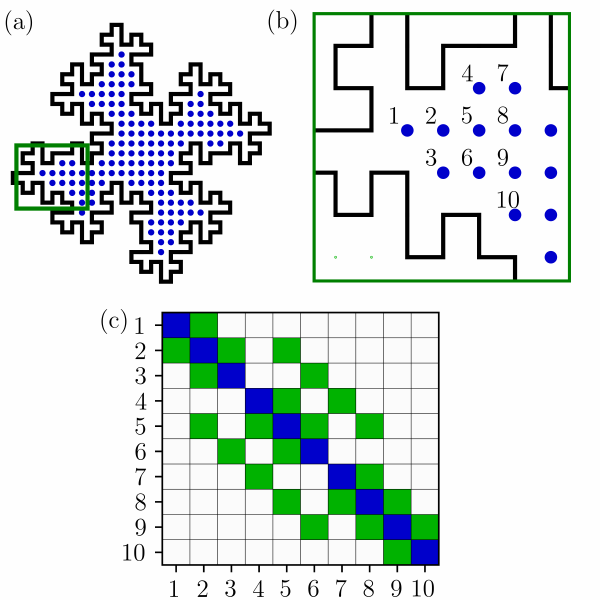} 
 \caption{The classification of the lattice points that are internal to the square Koch curve. (a) For the square Koch curve (solid black line) the blue solid dots represent internal lattice points. Points that are on the boundary or outside the fractal are not shown. The green box indicates the region that is detailed in panel (b) of this figure; (b) Assuming that the lattice is traversed column-by-column from the upper left corner, the values of the classification array corresponding internal points using the convention detailed in the main text are presented; (c) The structure of the coefficient matrix represented by the left-hand-side of Eq.~\eqref{eq:Helmholtz-FD}, that is, the finite difference approximation to the negative Laplacian $-\nabla^2$. Here blue squares represent the value $4/h^2$, the green squares represent the value $-1/h^2$ while the white squares represent the zero elements.}
 \label{Fig:classification}.
\end{figure}

To fill the whole classification array, we start from the upper left corner of the lattice, a point that corresponds to lattice point $\vec{r}_{11}$, and traverse the lattice column-by-column~\footnote{Alternatively, you can choose to traverse the lattice row-by-row without any changes to the eigenmodes that you calculate in the end. For speed purposes, your best option is to traverse the array in the way it is linearly stored in memory.}. For each lattice point, one of the methods outlined above (or others) is used to determine if the lattice point is inside or outside of the square Koch curve. For the calculations that we present in this paper, we used the winding number algorithm. If the lattice point is outside the square Koch curve, we set the value to $-1$ (or any other negative value). On the other hand, for lattice points that are classified as being inside, the classification array is given a strictly positive integer value. The classification value of the first internal point that we encounter is set to $1$, the second one to $2$, and so on. This way of labeling the internal lattice points will be convenient when we later set up the eigensystem (see the next subsection). When the lattice is traversed column-by-column starting from the upper left corner, as we have assumed here, the classification of the first internal lattice points is detailed in Fig.~\ref{Fig:classification}(b).

\subsection{Constructing the eigensystem}
\label{sec:3.4}

Equation~\eqref{eq:Helmholtz-FD} is the starting point for setting up the eigensystem that determines the eigenmodes and corresponding eigenfrequencies of the drum. However, we want the eigenfrequencies that we calculate to be independent of the width and height, $L$, of the square from which the square Koch drum was generated. Therefore, we multiply both sides of Eq.~\eqref{eq:Helmholtz-FD} by $L^2$ and define the \emph{dimensionless eigenfrequency}
\begin{align}
 \Omega
 &= \frac{\omega}{c} L,
\end{align}
of the square Koch drum. From the equation that is obtained in this way we construct the eigensystem $A\vec{v}=\lambda\vec{v}$.  
Here $A$ is the coefficient matrix representing the finite difference approximation to the negative of the Laplacian (times $L^2$), $\vec{v}$ is the eigenvector, and $\lambda=\Omega^2$ is the corresponding eigenvalue. First, one needs to adopt a storage convention that maps onto a vector the set of the matrix elements $U_{mn}$ that correspond to internal lattice points. We adopt the convention
\begin{align}
 \vec{v} &= \left( U_{m_1 n_1},\, U_{m_2 n_2},\, U_{m_3 n_3},\, \cdots \right)^T,
 \label{eq:Storage-convention}     
\end{align}
where the index pair $m_p n_p$ that appears as subscripts is defined from the lattice point classification matrix $C$ by $C(m_p,n_p)~=~p$ with $p$ a positive integer [$p\in {\mathbb N}^+$]. In other words, the $p$'th element of the eigenvector $\vec{v}$ corresponds to the lattice point located at position $(m_p, n_p)$. 

With this convention, and the use of the classification matrix $C$, the coefficient matrix $A$ can be constructed in the following way. First, all elements of the matrix $A$ are initialized to zero [$A=0$]. Then one loops over all lattice points (here in a column-by-column manner), $m=1,2,\ldots,N_\ell+1$ and $n=1,2,\ldots, N_\ell+1$. If a lattice point is outside or on the boundary of the square Koch curve, do nothing, and go on to the next lattice point. On the other hand, if the point of lattice indices $(m,n)$ is an internal point $i~=~C(m,n)~>~0$, the diagonal element of the coefficient matrix is set to $A_{ii}~=~ 4L^2/\delta_\ell^2~=~4^{\ell+1}$ [see Eq.~\eqref{eq:Helmholtz-FD}$\times L^2$] where we have used $h~=~\delta_\ell$ for the square Koch curve at generation level $\ell$. This value of $A_{ii}$ is indicated by the blue color in Fig.~\ref{Fig:classification}(c). Next, the potential coupling to its four nearest-neighboring lattice points is taken into account. This is done by subsequently considering the points that are located to the right and the left of the lattice point $(m,n)$, that is, points labeled $j=C(m+1,n)$ and $j=C(m-1,n)$, and the lattice points just above and below $(m,n)$ that are labeled $j=C(m,n+1)$ and $j=C(m,n-1)$. For each of the points that are nearest-neighbors to lattice point $(m,n)$ and also are internal lattice points so that $j>0$, one sets $A_{ij}= -L^2/\delta_\ell^2=-4^\ell$~[see Eq.~\eqref{eq:Helmholtz-FD}$\times L^2$]. Such elements are indicated by the green color in Fig.~\ref{Fig:classification}(c). In the same figure, the white color indicates vanishing (zero value) matrix elements. After completing the loop over the whole lattice, the coefficient matrix $A$ is filled and the eigenmodes and eigenvalues can be computed. One should note that the coefficient matrix $A$ is  symmetric and positive definite. Hence, the eigenvalues are real and the eigenvectors can be chosen to be real; this is required for the physical quantities frequency and displacement.

In passing, it should be noted that the matrix $A$ has dimension $M_\ell \times M_\ell$ where a good approximation for $M_\ell$ is $\left\lfloor (N_\ell  + 1 )^2  L^2/L_\ell^2 \right\rceil$. Furthermore, the majority of the elements of this matrix are zero, so it is a \emph{sparse matrix}. Taking advantage of the sparsity of the coefficient matrix $A$ is particularly important (to reduce memory requirements) if one wants to handle higher generation levels $\ell$. Since each row of the matrix $A$ can have at most $5$ non-zero elements, a lower bound on its sparsity~\footnote{The \emph{sparsity} of a matrix is defined as the ratio of the number of zero elements to the total number of elements of the matrix.} is $1 - 5/M_\ell$.


\subsection{Solving the eigensystem}
\label{sec:3.5}

 If the matrix $A$ is stored as a dense matrix~\footnote{This means that all elements of the matrix are stored, also the zero elements.}, the eigensystem is best solved by the routines \textsl{ssyev/dsyev} from the high-performance LAPACK-library~\cite{Lapack}. If instead the popular programming languages Python or \texttt{C++} are used, the Python modules NumPy/ScyPy~\cite{Book:Johansson2018,NumPy,SciPy} or the library Armadillo~\cite{Armadillo} will provide the same capabilities, while Matlab has an eigensolver directly built into the language. Internally, all these approaches use the LAPACK library. On the other hand, if you should opt for storing the coefficient matrix $A$ as a sparse matrix, ARPACK~\cite{Arpack} is the workhorse eigensolver library and both SciPy and Armadillo have wrappers to this library. Furthermore, Matlab handles sparse matrices as part of the language. It should be mentioned that ARPACK also has the option of calculating a given number of the lowest eigenvalues and corresponding eigenvectors. This option can be significantly faster than calculating the full set of eigenvalues and eigenvectors.

Independently of how the eigensystem is solved, the result is a set of eigenvalues $\{ \lambda_\nu\}$ and the corresponding set of eigenvectors $\{ \vec{v}_\nu \}$ (with $\nu=1,2,\ldots$). Typically the calculated eigenvectors $\vec{v}_\nu$ are calculated using a given normalization; for instance, if LAPACK is used for the calculation, the eigenvectors are normalized to have unit $L_2$-norms.

The calculated eigenvectors $\vec{v}_\nu$ cannot be visualized directly. Instead, they have to be mapped back onto the lattice that was initially defined and assumed in setting up the eigensystem (a mapping from a vector to a portion of a matrix). To this end, an eigenmode matrix $E_\nu$ is allocated to have the same dimensions as the lattice and the classification matrix $C$. By performing a (column-by-column) double loop over the elements $C(m,n)$ of the classification matrix~\footnote{This means that the inner loop is $m=1,2,\ldots,N_4+1$ while the outer loop is $n=1,2,\ldots,N_4+1$.}, such a vector-to-matrix mapping can be achieved by using how the classification matrix was defined~[see Sec.~\ref{sec:3.3}]. For points of the lattice $(m,n)$ that are not internal to the square Koch drum, indicated by  $C(m,n)\leq 0$, we put $E_\nu(m,n)=0$, \textit{i.e.} vanishing vertical displacement. However, for points of the lattice for which $C(m,n)>0$, we set $E_\nu(m,n)=v_\nu(i)$ where $i=C(m,n)$ is a positive integer~[see Sec.~\ref{sec:3.3} for details]. When the double-loop over $m$ and $n$ finishes, the vector-to-matrix mapping is completed and now the eigenmode can be visualized by generating a contour plot of the eigenmode matrix $E_\nu$ and on it  superposing the boundary of the square Koch curve assumed in calculating the eigenmodes. In this way, we obtained the eigenmodes that will be presented below (in Figs.~\ref{Fig:fractal_modes} and \ref{Fig:fractal_modes-HighOrder}.)




\section{\label{sec:4}Results and discussion}

The previous section detailed how to set up and solve the eigensystem $A\vec{v} = \lambda \vec{v}$ that determines the eigenmodes and eigenfrequencies of the square Koch drum. Here we will present and discuss the results that can be obtained by doing so. It will be assumed that the boundary of the square Koch drum is generated at level $\ell=4$~\footnote{Also satisfactory results can be obtained using the value $\ell=3$.}. This value of $\ell$ is high enough that the square Koch curve displays sufficient details without the resulting eigensystem taking too long to solve or requiring more memory than can be stored on a typical student laptop. For level $\ell=4$ the discretization interval is $\delta_4=L/4^4=L/256$~[Eq.~\eqref{eq:discretization-inteval-ell}], and the width of the square Koch drum is $L_4\approx 1.664 L$~[Eq.~\eqref{eq:L_over_Leff}].
Furthermore, with these values, or from Eq.~\eqref{eq:No-of-intervals}, it follows that the linear size of the quadratic lattice is $N_4+1=427$. Out of the $(N_4+1)^2=\num{182329}$ points, \num{16384} lattice points are boundary points, while there are $M_4=\num{57345}$ internal lattice points for the square Koch drum ($\ell=4$). Therefore, the size of the eigensystem is $M_4\times M_4$. Using single-precision floating points, dense storage of the coefficient matrix of the eigensystem will require about \SI{12.25}{Gb} of memory. Since the sparsity of the matrix is over \SI{99.9}{\percent}, only a fraction of this storage is required if sparse matrix storage is used. It should be mentioned that the students do not typically have sufficient memory on their laptops for dense matrix storage when $\ell\geq 4$; however, if they are using sparse storage, they are not expected to face this problem, until $\ell \geq 6$. 

\begin{figure}[htp!]
 \centering
 \includegraphics[width=0.99 \columnwidth]{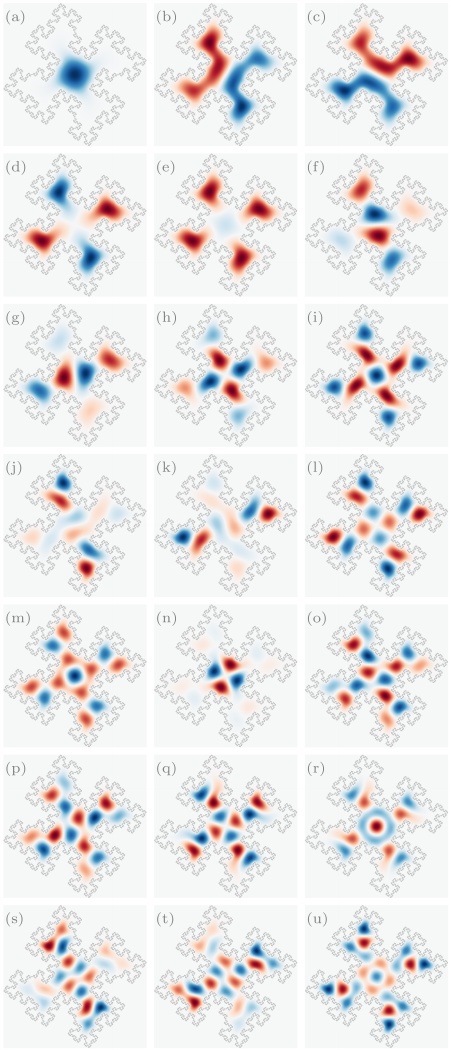}
 \caption{(Color online) The lowest eigenmodes $E_0$--$E_{20}$ of the square Koch drum at fractal generation level $\ell=4$ that correspond the lowest eigenfrequencies which are listed in Table~\ref{Tab:eigen-frequences}. These modes were obtained by solving the eigensystem as explained in Sec.~\ref{Sec:Implementation} assuming the discretization interval $\delta_4=L/4^4=L/256$. The blue and red colors represent negative and positive values for the vertical displacement, respectively. }
 \label{Fig:fractal_modes}
\end{figure}

\begin{figure}[htp!]
 \centering
 \includegraphics[width=0.33 \columnwidth,angle=90]{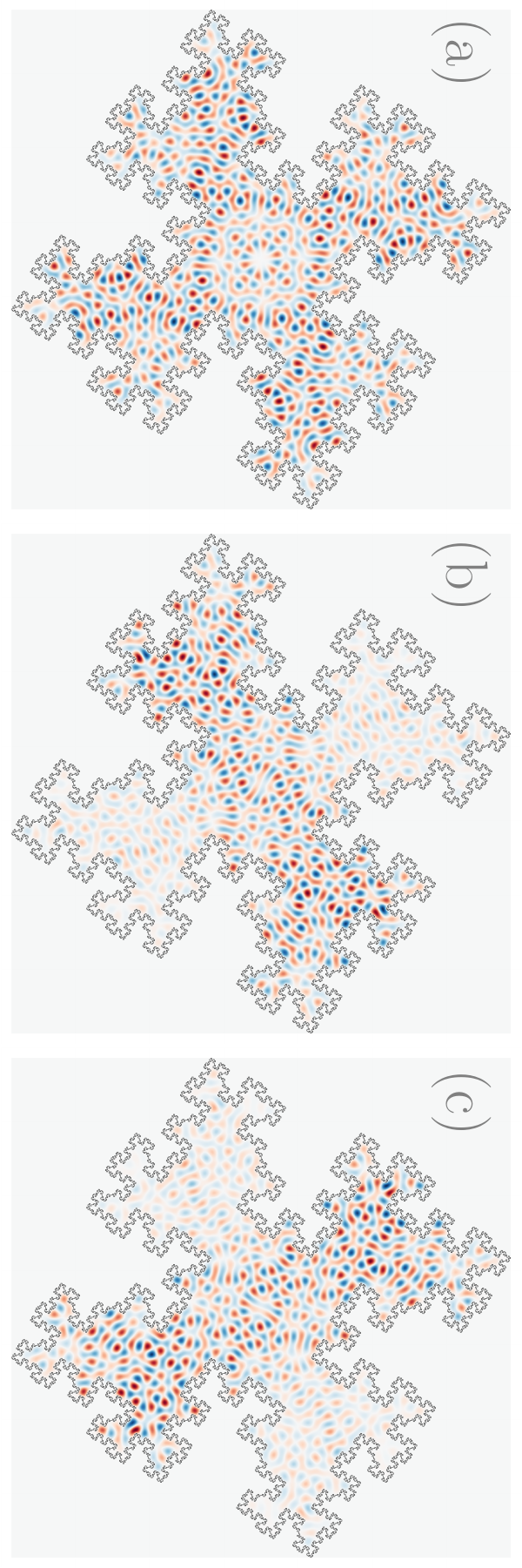}
 \caption{(Color online) The structure of the eigenmodes $E_{1113}$--$E_{1115}$ that correspond to the eigenfrequencies $\Omega_{1113}=136.3287$, and $\Omega_{1114}=\Omega_{1115}=136.3656$. The remaining parameters are like in Fig.~\ref{Fig:fractal_modes}. }
   \label{Fig:fractal_modes-HighOrder}
\end{figure}

For $\ell=4$ the eigensystem was constructed using sparse matrix storage and solved as outlined in Sec.~\ref{sec:3}. The calculation of the first \num{21} eigenmodes of the square Koch drum took only a few minutes on a typical desktop computer; the most time-consuming steps of the calculation were (\textit{i})~to obtain the classification of the lattice points, needed for the system setup, and (\textit{ii})~to solve the eigensystem. In this way we obtained the eigenmodes presented in Figs.~\ref{Fig:fractal_modes} and \ref{Fig:fractal_modes-HighOrder}. Here the calculated eigenvectors were mapped back onto the eigenmode matrix $E_\nu$ and contour plots of these modes, with the boundary of the square Koch drum superimposed, were produced to visualize the calculated modes [see Sec.~\ref{sec:3.5} for details]. 


\smallskip
Figure~\ref{Fig:fractal_modes}(a) presents the fundamental eigenmode of the square Koch drum (at level $\ell=4$). It is found that the vertical displacement of this mode is concentrated around the center of the square Koch drum and the displacement values all have the same sign; therefore, no nodal lines exist for the fundamental mode, as expected from the Courant nodal domain theorem~\cite{Book:Courant1989}. This feature is similar to the fundamental mode of the non-fractal square drum~[Fig.~\ref{Fig:Koch-fractal}(a)]~\cite{Book:Butkov1973,Book:Wong2013-NormalDrum}. The corresponding dimensionless eigenfrequency is $\Omega_0=9.4299$, a value that should be compared to the fundamental frequency of the square drum which is $\widehat{\Omega}_0=\sqrt{2}\pi=4.4429$~\cite{Sapoval1991,Book:Butkov1973,Book:Wong2013-NormalDrum}. Therefore, the ratio of these two fundamental frequencies is $\Omega_0/\widehat{\Omega}_0=2.1225$, a ratio that Sapoval~{\textit{et al.}} reported to be \num{2.100}~\cite{Sapoval1991}. Reducing the generation level to $\ell=3$, as assumed in the experiments by Sapoval~\textit{et al.}, resulted in a reduced ratio $\Omega_0/\widehat{\Omega}$ that still remained slightly higher than the experimental value. However, visually comparing the fundamental eigenmode in Fig.~\ref{Fig:fractal_modes}(a) to the fundamental mode depicted in Fig.~4(a) of Ref.~\onlinecite{Sapoval1991} shows good agreement.

With regards to the eigenmodes $E_1$ and $E_2$, seen in Figs.~\ref{Fig:fractal_modes}~(b) and \ref{Fig:fractal_modes}(c), we numerically find that $\Omega_1$ equals $\Omega_2$ to \num{10} decimal places~[Table~\ref{Tab:eigen-frequences}], which we interpret as a sign of \emph{degeneracy}.
The number of different eigenmodes corresponding to a particular eigenfrequency is known as the \emph{degree of degeneracy}. It should be recalled that the first excited states of a square drum are also degenerate with a degree of degeneracy of two~\cite{Book:Butkov1973}.

The following two eigenmodes, $E_3$ and $E_4$, are non-degenerate and their structures are presented in Figs.~\ref{Fig:fractal_modes}(d) and \ref{Fig:fractal_modes}(e). For both these modes, the displacement is mainly in the four ``wings'' of the square Koch drum, while, for each mode, the displacement at the center of the drum is significantly lower. Hence, one observes four well-defined regions for which the displacement is significant. For mode $E_4$, the displacement in these regions has the same sign, while for mode $E_3$, two diagonally placed regions have positive displacement while the other two have negative displacement. The reason the $E_3$ eigenmode does not have a rotated, degenerate eigenmode is discussed later in this section and can be explained on the basis of group theory.
If we compare the eigenmodes $E_1$--$E_4$ from Figs.~\ref{Fig:fractal_modes}(b)--(e) (and their eigenfrequencies), to the corresponding modes shown by Fig.~5 in Sapoval~{\textit{et al.}}~\cite{Sapoval1991}, good qualitative agreement is found. 
It is remarked that the experimental displacement pattern presented in Fig.~\ref{Fig:sapoval}(b) can be obtained by a linear combination of the modes $E_1$--$E_4$, as was explained in Ref.~\onlinecite{Sapoval1991}. 

\begin{table} 
 \caption{\label{Tab:eigen-frequences}
 The eigenfrequencies associated with the eigenmodes of the square Koch drum ($\ell=4$) depicted in Figs.~\ref{Fig:fractal_modes}. The columns of the table present the mode index $\nu$, the dimensionless eigenfrequency $\Omega_\nu$ and the degree of degeneracy $g_\nu$, both for the square Koch drum, and finally the ratio $\Omega_\nu/ \widehat{\Omega}_0$ where $\widehat{\Omega}_0=\sqrt{2}\pi$ is the dimensionless fundamental eigenfrequency of the corresponding classic square drum.}
 \begin{ruledtabular}
 \begin{tabular}{crcc}
  $\nu$\; & $\Omega_\nu$\;\;\; & $g_\nu$ & $\Omega_\nu/ \widehat{\Omega}_0$
   \\
   \hline
   \\
   0	 &    9.4299    &  1  &   2.1225
   \\
   1	 &   14.1469    &  2  &   3.1842 
   \\   
   2	 &   14.1469    &  2  &   3.1842 
   \\
   3	 &   14.4199    &  1  &   3.2456 
   \\
   4	 &   14.4969    &  1  &   3.2629
   \\
   5	 &   15.0824    &  2  &   3.3947 
   \\
   6	 &   15.0824    &  2  &   3.3947 
   \\
   7	 &   17.6559    &  1  &   3.9740 
   \\
   8	 &   18.9114    &  1  &   4.2565 
   \\
   9	 &   19.4563    &  2  &   4.3792 
   \\
   10 & 	19.4563    &  2  &   4.3792
   \\
   11 & 	20.0210    &  1  &   4.5063 
   \\
   12 & 	20.5972    &  1  &   4.6360 
   \\
   13 & 	21.3443    &  1  &   4.8041 
   \\
   14 & 	21.6361    &  2  &   4.8698 
   \\
   15 & 	21.6361    &  2  &   4.8698
   \\
   16 & 	23.3219    &  1  &   5.2492 
   \\
   17 & 	23.5807    &  1  &   5.3075 
   \\
   18 & 	24.8755    &  2  &   5.5989 
   \\
   19 & 	24.8755    &  2  &   5.5989 
   \\
   20 & 	25.7253    &  1  &   5.7902
  \end{tabular}
 \end{ruledtabular}
\end{table}
Figures~\ref{Fig:fractal_modes}(f)--(u) present the structure of the modes $E_\nu$ for $\nu=5$--$20$ and their corresponding eigenfrequencies are given in Table~\ref{Tab:eigen-frequences}. Several of these modes are degenerate, like the modes that correspond to mode indices  $\nu=5,6$; $\nu=9,10$; $\nu=14,15$ and $\nu=18,19$ [see Table~\ref{Tab:eigen-frequences}]. Moreover, and as expected, one finds that the spatial complexity of the modes increases with the mode index. It is hard not to appreciate the esthetic beauty of some of these higher-order modes depicted in Fig.~\ref{Fig:fractal_modes}. Many students found motivation in producing, on their own account, such appealing results.

One may also wonder what some of the much higher-order modes of the square Koch drum look like. To this end, Fig.~\ref{Fig:fractal_modes-HighOrder} presents the modes $E_{1113}$--$E_{1115}$. The associated eigenfrequencies are given in the figure caption. The mode structure is rather complex, as expected, and $E_{1114}$ and $E_{1115}$ are, in fact, degenerate modes. 


\medskip

We now turn to the symmetry properties of the eigenmodes presented in Figs.~\ref{Fig:fractal_modes} and \ref{Fig:fractal_modes-HighOrder}. These properties are determined by the symmetries of the eigenproblem~\eqref{eq:Helmholtz}. The square Koch curve~[Fig.~\ref{Fig:Koch-fractal}(c)] is invariant with respect to in-plane rotations of \ang{90} about the center of the drum (for any value of $\ell$). Since the Helmholtz equation~\eqref{eq:Helmholtz-A} is rotationally invariant, the full solution to~\eqref{eq:Helmholtz} displays in-plane \ang{90}-rotational symmetry. The consequence for the eigenmodes of this symmetry is typically studied using group theory~\cite{Book:Arfken2012,Book:Hamermesh1989}. The useful result to note from such theory is that when a symmetry operation of the problem is applied to one of its eigenmodes, the result will be a linear combination of the eigenmodes corresponding to the same eigenvalue~\cite{Book:Hamermesh1989}. This has the consequence that non-degenerate eigenmodes of the square Koch drum should, up to a constant, be \ang{90}-rotational symmetric about their center point. For a $g_\nu=2$ degenerate eigenmode, the prediction is that its in-plane rotation of \ang{90} about its center should, due to the orthogonality of the eigenmodes, result in a constant times the other eigenmode that corresponds to the same eigenvalue. Close inspection of the modes in Figs.~\ref{Fig:fractal_modes} and \ref{Fig:fractal_modes-HighOrder} reveals that the expected symmetry properties are indeed present in the calculated eigenmodes.

In total $11$ of the $21$ eigenmodes of the square Koch drum presented in Fig.~\ref{Fig:fractal_modes} are non-degenerate~[Table~\ref{Tab:eigen-frequences}]. The dimensionless eigenfrequencies of the (non-fractal) square drum are $\sqrt{m^2+n^2}\pi$ with $m,n=1,2,\ldots$~\cite{Book:Butkov1973,SquareDrum2D}. Among the $21$ first eigenmodes of the square drum, only $4$ modes are non-degenerate. The lower number of degenerate eigenmodes found for the square Koch drum as compared to the corresponding non-fractal square drum is due to the latter drum having a higher degree of symmetry. The classic square drum is also symmetric with respect to reflections about the first (horizontal) and second (vertical) axis~[Fig.~\ref{Fig:fractal-generator}(a)] and with respect to the $\pm \ang{45}$ diagonals. These symmetries are \textit{not} present for the square Koch drum. For this reason, some of the degeneracy that is present in the classic square drum is lifted for the corresponding square Koch drum. Additional symmetry in the shape of the drum increases the fraction of eigenmodes that are degenerate, at least, this is the case for the drums that we considered.

\section{\label{sec:5} Conclusions}

The numerical experiment described in this paper provided students with a better understanding of the vibrational properties of fractal or extremely irregular structures. Important topics include the vibrations of fractal drums, their eigenfrequencies and corresponding eigenstates. Optionally, one could extend the study to include the density of states in order to examine the Weyl-Berry conjecture.

The numerical experiment allows students to construct a fractal drum, calculate its eigenmodes, and visualize the vibrational modes. The students can change boundary conditions, vary certain dimensions, and observe the results. The assignment may be integrated into a computational physics class. Understanding students’ concerns when solving a numerical problem allows the teacher to be more effective and help all their students take full advantage of the educational resources at their disposal. The ideal group size for conducting the proposed activities is two students to allow for discussions between them. Furthermore, this problem will expose students to eigenvalue problems which probably are larger than what they have faced during their studies. In order to solve it, they have to generate the fractal structure and must learn to master how to map an unorganized portion of a matrix of unknowns into a vector (required by the eigensolver), and to define the coefficient matrix that is associated with it. Since this matrix is quite sparse, the use of eigensolvers for sparse matrices will typically become a topic of interest. Last but not least, our experience in presenting/supervising this computational student project several times is that the students tend to enjoy it. Students typically find the project challenging but are still motivated to solve the problem; they are fascinated by the beauty of some of the eigenmodes of the square Koch drum. The hope is that others can benefit from our experience with this numerical student experiment. 

Many of the tasks in this numerical experiment presented students with novel challenges. For example, students working on the classification of whether lattice points are inside or outside the fractal boundary struggled with finding an efficient solution. 

Since some of the tasks in this work involve very large arrays such as the coefficient matrix, every portion of the code must be optimized to yield a solution within a realistic time span. Students reported that while constructing and solving the eigensystem was relatively simple, optimizing this process was more challenging. Furthermore, they also reported that the scope and difficulty of the tasks of this numerical experiment improved their confidence in their own coding abilities for the purpose of both scientific numerical modeling and software engineering.

To assist instructors considering applying the ``fractal drum'' project discussed in this paper, the formulation of the project as we used it in our course,  including the step-by-step instructions for the students, is available as supplementary material in Ref.~\onlinecite{SupMat}.

\begin{acknowledgments}
V.P.S. acknowledges the Research Council of Norway through its Center of Excellence Funding Scheme, Project No.~262644 PoreLab, for allowing her the use of PoreLab's facilities, and I.S. thanks Dr.~J.O.~Fj{\ae}restad for fruitful discussions on group theory. The authors gratefully acknowledge the anonymous referees and the editor whose constructive comments improved this paper.
\end{acknowledgments}

%
\bibliographystyle{apsrev4-2}

\begin{thebibliography}{29}%
\makeatletter
\providecommand \@ifxundefined [1]{%
 \@ifx{#1\undefined}
}%
\providecommand \@ifnum [1]{%
 \ifnum #1\expandafter \@firstoftwo
 \else \expandafter \@secondoftwo
 \fi
}%
\providecommand \@ifx [1]{%
 \ifx #1\expandafter \@firstoftwo
 \else \expandafter \@secondoftwo
 \fi
}%
\providecommand \natexlab [1]{#1}%
\providecommand \enquote  [1]{``#1''}%
\providecommand \bibnamefont  [1]{#1}%
\providecommand \bibfnamefont [1]{#1}%
\providecommand \citenamefont [1]{#1}%
\providecommand \href@noop [0]{\@secondoftwo}%
\providecommand \href [0]{\begingroup \@sanitize@url \@href}%
\providecommand \@href[1]{\@@startlink{#1}\@@href}%
\providecommand \@@href[1]{\endgroup#1\@@endlink}%
\providecommand \@sanitize@url [0]{\catcode `\\12\catcode `\$12\catcode
  `\&12\catcode `\#12\catcode `\^12\catcode `\_12\catcode `\%12\relax}%
\providecommand \@@startlink[1]{}%
\providecommand \@@endlink[0]{}%
\providecommand \url  [0]{\begingroup\@sanitize@url \@url }%
\providecommand \@url [1]{\endgroup\@href {#1}{\urlprefix }}%
\providecommand \urlprefix  [0]{URL }%
\providecommand \Eprint [0]{\href }%
\providecommand \doibase [0]{https://doi.org/}%
\providecommand \selectlanguage [0]{\@gobble}%
\providecommand \bibinfo  [0]{\@secondoftwo}%
\providecommand \bibfield  [0]{\@secondoftwo}%
\providecommand \translation [1]{[#1]}%
\providecommand \BibitemOpen [0]{}%
\providecommand \bibitemStop [0]{}%
\providecommand \bibitemNoStop [0]{.\EOS\space}%
\providecommand \EOS [0]{\spacefactor3000\relax}%
\providecommand \BibitemShut  [1]{\csname bibitem#1\endcsname}%
\let\auto@bib@innerbib\@empty
\bibitem [{\citenamefont {Kac}(1966)}]{Kac1966}%
  \BibitemOpen
  \bibfield  {author} {\bibinfo {author} {\bibfnamefont {M.}~\bibnamefont
  {Kac}},\ }\href {http://www.jstor.org/stable/2313748} {\bibfield  {journal}
  {\bibinfo  {journal} {Am. Math. Mon.}\ }\textbf {\bibinfo {volume} {73}},\
  \bibinfo {pages} {1} (\bibinfo {year} {1966})}\BibitemShut {NoStop}%
\bibitem [{\citenamefont {Feder}(1988)}]{Book:Feder1988}%
  \BibitemOpen
  \bibfield  {author} {\bibinfo {author} {\bibfnamefont {J.}~\bibnamefont
  {Feder}},\ }\href@noop {} {\emph {\bibinfo {title} {Fractals}}}\ (\bibinfo
  {publisher} {Plenum Press},\ \bibinfo {address} {New York},\ \bibinfo {year}
  {1988})\BibitemShut {NoStop}%
\bibitem [{\citenamefont {Sapoval}\ \emph {et~al.}(1991)\citenamefont
  {Sapoval}, \citenamefont {Gobron},\ and\ \citenamefont
  {Margolina}}]{Sapoval1991}%
  \BibitemOpen
  \bibfield  {author} {\bibinfo {author} {\bibfnamefont {B.}~\bibnamefont
  {Sapoval}}, \bibinfo {author} {\bibfnamefont {T.}~\bibnamefont {Gobron}},\
  and\ \bibinfo {author} {\bibfnamefont {A.}~\bibnamefont {Margolina}},\ }\href
  {http://web.phys.ntnu.no/~ingves/Teaching/TFY4235/Exam/Download/Sapoval1991.pdf}
  {\bibfield  {journal} {\bibinfo  {journal} {Phys. Rev. Lett.}\ }\textbf
  {\bibinfo {volume} {67}},\ \bibinfo {pages} {2974} (\bibinfo {year}
  {1991})}\BibitemShut {NoStop}%
\bibitem [{\citenamefont {Butkov}(1973)}]{Book:Butkov1973}%
  \BibitemOpen
  \bibfield  {author} {\bibinfo {author} {\bibfnamefont {E.}~\bibnamefont
  {Butkov}},\ }\href@noop {} {\emph {\bibinfo {title} {Mathematical Physics}}}\
  (\bibinfo  {publisher} {Addison-Wesley Publishing Company},\ \bibinfo
  {address} {Reading, MA},\ \bibinfo {year} {1973})\ pp.\ \bibinfo {pages}
  {313--325}\BibitemShut {NoStop}%
\bibitem [{\citenamefont {Wong}(2013{\natexlab{a}})}]{Book:Wong2013}%
  \BibitemOpen
  \bibfield  {author} {\bibinfo {author} {\bibfnamefont {C.~W.}\ \bibnamefont
  {Wong}},\ }\href@noop {} {\emph {\bibinfo {title} {Introduction to
  Mathematical Physics: Methods \& Concepts}}},\ \bibinfo {edition} {2nd}\ ed.\
  (\bibinfo  {publisher} {Oxford University Press},\ \bibinfo {year} {2013})\
  pp.\ \bibinfo {pages} {118--119}\BibitemShut {NoStop}%
\bibitem [{\citenamefont {Sauer}(2012)}]{Book:Sauer2012}%
  \BibitemOpen
  \bibfield  {author} {\bibinfo {author} {\bibfnamefont {T.}~\bibnamefont
  {Sauer}},\ }\href@noop {} {\emph {\bibinfo {title} {Numerical Analysis}}},\
  \bibinfo {edition} {2nd}\ ed.\ (\bibinfo  {publisher} {Pearson},\ \bibinfo
  {address} {Boston},\ \bibinfo {year} {2012})\ \bibinfo {note}
  {{S}ec.~8.3.1}\BibitemShut {NoStop}%
\bibitem [{\citenamefont {Abramowitz}\ and\ \citenamefont
  {Stegun}(1964)}]{Book:Abramowitz1964}%
  \BibitemOpen
  \bibfield  {author} {\bibinfo {author} {\bibfnamefont {M.}~\bibnamefont
  {Abramowitz}}\ and\ \bibinfo {author} {\bibfnamefont {I.}~\bibnamefont
  {Stegun}},\ }\href
  {http://www.ebook.de/de/product/1675514/handbook_of_mathematical_functions.html}
  {\emph {\bibinfo {title} {Handbook of Mathematical Functions with Formulas,
  Graphs, and Mathematical Tables}}}\ (\bibinfo  {publisher} {Dover
  Publications},\ \bibinfo {year} {1964})\ \bibinfo {note}
  {{S}ec.~25.3.30}\BibitemShut {NoStop}%
\bibitem [{\citenamefont {Hughes}\ \emph {et~al.}(2014)\citenamefont {Hughes},
  \citenamefont {van Dam}, \citenamefont {McGuire}, \citenamefont {Sklar},
  \citenamefont {Foley}, \citenamefont {Feiner},\ and\ \citenamefont
  {Akeley}}]{Book:Hughes2013}%
  \BibitemOpen
  \bibfield  {author} {\bibinfo {author} {\bibfnamefont {J.}~\bibnamefont
  {Hughes}}, \bibinfo {author} {\bibfnamefont {A.}~\bibnamefont {van Dam}},
  \bibinfo {author} {\bibfnamefont {M.}~\bibnamefont {McGuire}}, \bibinfo
  {author} {\bibfnamefont {D.}~\bibnamefont {Sklar}}, \bibinfo {author}
  {\bibfnamefont {J.}~\bibnamefont {Foley}}, \bibinfo {author} {\bibfnamefont
  {S.}~\bibnamefont {Feiner}},\ and\ \bibinfo {author} {\bibfnamefont
  {K.}~\bibnamefont {Akeley}},\ }\href@noop {} {\emph {\bibinfo {title}
  {Computer Graphics: Principles and Practice}}},\ \bibinfo {edition} {3rd}\
  ed.\ (\bibinfo  {publisher} {Addison-Wesley},\ \bibinfo {address} {Upper
  Saddle River, NJ},\ \bibinfo {year} {2014})\ \bibinfo {note}
  {{S}ec.~7.10.1}\BibitemShut {NoStop}%
\bibitem [{\citenamefont {Hormann}\ and\ \citenamefont
  {Agathos}(2001)}]{HORMANN2001131}%
  \BibitemOpen
  \bibfield  {author} {\bibinfo {author} {\bibfnamefont {K.}~\bibnamefont
  {Hormann}}\ and\ \bibinfo {author} {\bibfnamefont {A.}~\bibnamefont
  {Agathos}},\ }\href
  {https://doi.org/https://doi.org/10.1016/S0925-7721(01)00012-8} {\bibfield
  {journal} {\bibinfo  {journal} {Comp. Geom.}\ }\textbf {\bibinfo {volume}
  {20}},\ \bibinfo {pages} {131} (\bibinfo {year} {2001})}\BibitemShut
  {NoStop}%
\bibitem [{Wik()}]{Wikipedia-PIP}%
  \BibitemOpen
  \href@noop {} {}\bibinfo {note} {Wikipedia page on ``Point in Polygon``:
  \url{https://en.wikipedia.org/wiki/Point_in_polygon}}\BibitemShut {NoStop}%
\bibitem [{\citenamefont {Sutherland}\ \emph {et~al.}(1974)\citenamefont
  {Sutherland}, \citenamefont {Sproull},\ and\ \citenamefont
  {Schumacker}}]{Sutherland1974}%
  \BibitemOpen
  \bibfield  {author} {\bibinfo {author} {\bibfnamefont {I.~E.}\ \bibnamefont
  {Sutherland}}, \bibinfo {author} {\bibfnamefont {R.~F.}\ \bibnamefont
  {Sproull}},\ and\ \bibinfo {author} {\bibfnamefont {R.~A.}\ \bibnamefont
  {Schumacker}},\ }\href@noop {} {\bibfield  {journal} {\bibinfo  {journal}
  {ACM Comput. Surv.}\ }\textbf {\bibinfo {volume} {6}},\ \bibinfo {pages} {1}
  (\bibinfo {year} {1974})}\BibitemShut {NoStop}%
\bibitem [{\citenamefont {Narasimhan}(1985)}]{Narasimhan1985}%
  \BibitemOpen
  \bibfield  {author} {\bibinfo {author} {\bibfnamefont {R.}~\bibnamefont
  {Narasimhan}},\ }\bibinfo {title} {The winding number and the residue
  theorem},\ in\ \href {https://doi.org/10.1007/978-1-4757-1106-6_3} {\emph
  {\bibinfo {booktitle} {Complex Analysis in one Variable}}}\ (\bibinfo
  {publisher} {Birkh{\"a}user},\ \bibinfo {address} {Boston, MA},\ \bibinfo
  {year} {1985})\ pp.\ \bibinfo {pages} {70--88}\BibitemShut {NoStop}%
\bibitem [{Note1()}]{Note1}%
  \BibitemOpen
  \bibinfo {note} {Alternatively, you can choose to traverse the lattice
  row-by-row without any changes to the eigenmodes that you calculate in the
  end. For speed purposes, your best option is to traverse the array in the way
  it is linearly stored in memory.}\BibitemShut {Stop}%
\bibitem [{Note2()}]{Note2}%
  \BibitemOpen
  \bibinfo {note} {The \protect \emph {sparsity} of a matrix is defined as the
  ratio of the number of zero elements to the total number of elements of the
  matrix.}\BibitemShut {Stop}%
\bibitem [{Note3()}]{Note3}%
  \BibitemOpen
  \bibinfo {note} {This means that all elements of the matrix are stored, also
  the zero elements.}\BibitemShut {Stop}%
\bibitem [{\citenamefont {Anderson}\ \emph {et~al.}(1999)\citenamefont
  {Anderson}, \citenamefont {Bai}, \citenamefont {Bischof}, \citenamefont
  {Blackford}, \citenamefont {Dongarra}, \citenamefont {Croz}, \citenamefont
  {Greenbaum}, \citenamefont {Hammarling}, \citenamefont {McKenney},\ and\
  \citenamefont {Sorensen}}]{Lapack}%
  \BibitemOpen
  \bibfield  {author} {\bibinfo {author} {\bibfnamefont {E.}~\bibnamefont
  {Anderson}}, \bibinfo {author} {\bibfnamefont {Z.}~\bibnamefont {Bai}},
  \bibinfo {author} {\bibfnamefont {C.}~\bibnamefont {Bischof}}, \bibinfo
  {author} {\bibfnamefont {S.}~\bibnamefont {Blackford}}, \bibinfo {author}
  {\bibfnamefont {J.~D.~J.}\ \bibnamefont {Dongarra}}, \bibinfo {author}
  {\bibfnamefont {J.~D.}\ \bibnamefont {Croz}}, \bibinfo {author}
  {\bibfnamefont {A.}~\bibnamefont {Greenbaum}}, \bibinfo {author}
  {\bibfnamefont {S.}~\bibnamefont {Hammarling}}, \bibinfo {author}
  {\bibfnamefont {A.}~\bibnamefont {McKenney}},\ and\ \bibinfo {author}
  {\bibfnamefont {D.}~\bibnamefont {Sorensen}},\ }\href@noop {} {\emph
  {\bibinfo {title} {{LAPACK} {U}sers' {G}uide}}},\ \bibinfo {edition} {3rd}\
  ed.\ (\bibinfo  {publisher} {SIAM},\ \bibinfo {address} {Philadelphia,
  Pennsylvania, USA},\ \bibinfo {year} {1999})\BibitemShut {NoStop}%
\bibitem [{\citenamefont {Johansson}(2018)}]{Book:Johansson2018}%
  \BibitemOpen
  \bibfield  {author} {\bibinfo {author} {\bibfnamefont {R.}~\bibnamefont
  {Johansson}},\ }\href@noop {} {\emph {\bibinfo {title} {Numerical Python:
  Scientific Computing and Data Science Applications with Numpy, SciPy and
  Matplotlib}}},\ \bibinfo {edition} {2nd}\ ed.\ (\bibinfo  {publisher}
  {Apress},\ \bibinfo {year} {2018})\BibitemShut {NoStop}%
\bibitem [{Num()}]{NumPy}%
  \BibitemOpen
  \href@noop {} {}\bibinfo {note} {NumPy documentation, Version 1.24,
  \url{https://numpy.org/doc/1.24/}}\BibitemShut {NoStop}%
\bibitem [{Sci()}]{SciPy}%
  \BibitemOpen
  \href@noop {} {}\bibinfo {note} {SciPy documentation, Version 1.9.3,
  \url{https://docs.scipy.org/doc/}}\BibitemShut {NoStop}%
\bibitem [{\citenamefont {Sanderson}\ and\ \citenamefont
  {Curtin}(2016)}]{Armadillo}%
  \BibitemOpen
  \bibfield  {author} {\bibinfo {author} {\bibfnamefont {C.}~\bibnamefont
  {Sanderson}}\ and\ \bibinfo {author} {\bibfnamefont {R.}~\bibnamefont
  {Curtin}},\ }\href {https://arma.sourceforge.net/docs.html} {\bibfield
  {journal} {\bibinfo  {journal} {J. Open Source Softw.}\ }\textbf {\bibinfo
  {volume} {1}},\ \bibinfo {pages} {26} (\bibinfo {year} {2016})}\BibitemShut
  {NoStop}%
\bibitem [{\citenamefont {Lehoucq}\ \emph {et~al.}(1998)\citenamefont
  {Lehoucq}, \citenamefont {Sorensen},\ and\ \citenamefont {Yang}}]{Arpack}%
  \BibitemOpen
  \bibfield  {author} {\bibinfo {author} {\bibfnamefont {R.}~\bibnamefont
  {Lehoucq}}, \bibinfo {author} {\bibfnamefont {D.~C.}\ \bibnamefont
  {Sorensen}},\ and\ \bibinfo {author} {\bibfnamefont {C.}~\bibnamefont
  {Yang}},\ }\href@noop {} {\emph {\bibinfo {title} {{ARPACK} {U}sers {G}uide:
  Solution of Large-Scale Eigenvalue Problems with Implicitly Restarted Arnoldi
  Method}}}\ (\bibinfo  {publisher} {SIAM},\ \bibinfo {address}
  {Philadelphia},\ \bibinfo {year} {1998})\BibitemShut {NoStop}%
\bibitem [{Note4()}]{Note4}%
  \BibitemOpen
  \bibinfo {note} {This means that the inner loop is $m=1,2,\protect \ldots
  ,N_4+1$ while the outer loop is $n=1,2,\protect \ldots ,N_4+1$.}\BibitemShut
  {Stop}%
\bibitem [{Note5()}]{Note5}%
  \BibitemOpen
  \bibinfo {note} {Also satisfactory results can be obtained using the value
  $\ell =3$.}\BibitemShut {Stop}%
\bibitem [{\citenamefont {Courant}\ and\ \citenamefont
  {Hilbert}(1989)}]{Book:Courant1989}%
  \BibitemOpen
  \bibfield  {author} {\bibinfo {author} {\bibfnamefont {R.}~\bibnamefont
  {Courant}}\ and\ \bibinfo {author} {\bibfnamefont {D.}~\bibnamefont
  {Hilbert}},\ }\href@noop {} {\emph {\bibinfo {title} {Methods of Mathematical
  Physics}}},\ \bibinfo {edition} {2nd}\ ed.,\ Vol.~\bibinfo {volume} {1}\
  (\bibinfo  {publisher} {Wiley-VCH},\ \bibinfo {year} {1989})\ p.\ \bibinfo
  {pages} {454}\BibitemShut {NoStop}%
\bibitem [{\citenamefont
  {Wong}(2013{\natexlab{b}})}]{Book:Wong2013-NormalDrum}%
  \BibitemOpen
  \bibfield  {author} {\bibinfo {author} {\bibfnamefont {C.~W.}\ \bibnamefont
  {Wong}},\ }\href@noop {} {\emph {\bibinfo {title} {Introduction to
  Mathematical Physics: Methods \& Concepts}}},\ \bibinfo {edition} {2nd}\ ed.\
  (\bibinfo  {publisher} {Oxford University Press},\ \bibinfo {year} {2013})\
  pp.\ \bibinfo {pages} {225--226}\BibitemShut {NoStop}%
\bibitem [{\citenamefont {Arfken}\ \emph {et~al.}(2012)\citenamefont {Arfken},
  \citenamefont {Weber},\ and\ \citenamefont {Harris}}]{Book:Arfken2012}%
  \BibitemOpen
  \bibfield  {author} {\bibinfo {author} {\bibfnamefont {G.~B.}\ \bibnamefont
  {Arfken}}, \bibinfo {author} {\bibfnamefont {H.~J.}\ \bibnamefont {Weber}},\
  and\ \bibinfo {author} {\bibfnamefont {F.~E.}\ \bibnamefont {Harris}},\
  }\href@noop {} {\emph {\bibinfo {title} {Mathematical Methods for Physicists:
  A Comprehensive Guide}}},\ \bibinfo {edition} {7th}\ ed.\ (\bibinfo
  {publisher} {Academic Press},\ \bibinfo {address} {Amsterdam},\ \bibinfo
  {year} {2012})\ Chap.~\bibinfo {chapter} {17}\BibitemShut {NoStop}%
\bibitem [{\citenamefont {Hamermesh}(1989)}]{Book:Hamermesh1989}%
  \BibitemOpen
  \bibfield  {author} {\bibinfo {author} {\bibfnamefont {M.}~\bibnamefont
  {Hamermesh}},\ }\href@noop {} {\emph {\bibinfo {title} {Group Theory and Its
  Application to Physical Problems}}},\ Dover Books on Physics\ (\bibinfo
  {publisher} {Dover Publications},\ \bibinfo {year} {1989})\BibitemShut
  {NoStop}%
\bibitem [{Squ()}]{SquareDrum2D}%
  \BibitemOpen
  \href@noop {} {}\bibinfo {note} {Visualization of the eigenmodes of the
  square drum can be found at
  \url{https://www.compadre.org/PQP/quantum-theory/section13_1b.cfm}}\BibitemShut
  {NoStop}%
\bibitem [{Sup()}]{SupMat}%
  \BibitemOpen
  \href@noop {} {}\bibinfo {note} {Student instructions are available at [url
  inserted by AIPP].}\BibitemShut {Stop}%
\end{thebibliography}
%
%

\end{document}